\documentclass{emulateapj}

\shorttitle{EX Lupi Outburst in CO Fundamental band} 
\shortauthors{Goto et al.} 

\begin{document}

\title{Fundamental Vibrational Transition of CO during the Outburst
  of EX Lupi in 2008 \altaffilmark{1,2}}

\author{M. Goto,\altaffilmark{3}
Zs. Reg\'aly\altaffilmark{4},
C. P. Dullemond\altaffilmark{3},
M. van den Ancker\altaffilmark{5},
J. M. Brown\altaffilmark{6},
A. Carmona\altaffilmark{7},
K. Pontoppidan\altaffilmark{8},
P. \'Abrah\'am\altaffilmark{4},
G. A. Blake\altaffilmark{8},
D. Fedele\altaffilmark{3,9},
Th. Henning\altaffilmark{3},
A. Juh\'asz\altaffilmark{3},
\'A. K\'osp\'al\altaffilmark{10},
L. Mosoni\altaffilmark{3,4},
A. Sicilia-Aguilar\altaffilmark{3},
H. Terada\altaffilmark{11},
R. van Boekel\altaffilmark{3},
E. F. van Dishoeck\altaffilmark{8,10},
T. Usuda\altaffilmark{11}
}
\email{mgoto@mpia.de}

\altaffiltext{1}{Based on data collected in the course of CRIRES
                 Large program [179.C-0151] and DDT program
                 [281.C-4031(D,E,F)] at the {\it VLT} on Cerro
                 Paranal (Chile), which is operated by the
                 European Southern Observatory (ESO).}

\altaffiltext{2}{Based on data collected at Subaru Telescope, which is
                 operated by the National Astronomical
                 Observatory of Japan.}

\altaffiltext{3}{Max-Planck-Institut f\"ur Astronomie,
  K\"onigstuhl 17, D-69117 Heidelberg, Germany.}

\altaffiltext{4}{Konkoly Observatory, Konkoly
  Thege Mikl\'os 15-17, H-1121 Budapest, Hungary.}    

\altaffiltext{5}{ESO, Karl-Schwarzschild-Stra\ss e 2,
                 D-85748 Garching bei M\"unchen, Germany.} 

\altaffiltext{6}{Max-Planck-Institut f\"ur
                 extraterrestrische Physik, Giessenbachstra\ss e
                 D-85748 Garching bei M\"unchen, Germany.}

\altaffiltext{7}{ISDC Data Centre for Astrophysics \& Geneva
  Observatory, University of Geneva, chemin d'Ecogia 16, 1290
  Versoix, Switzerland.}

\altaffiltext{8}{California Institute of Technology Division of
                 Geological \& Planetary Sciences, MC~170-25
                 1200 E. California Blvd., Pasadena, CA 91125,
                 USA.}

\altaffiltext{9}{Johns Hopkins University, Department of Physics
  and Astronomy 3701 San Martin drive Baltimore, MD 21218 USA.}

\altaffiltext{10}{Leiden Observatory, Leiden University, P.O. Box 9513
	NL-2300 RA  Leiden, The Netherlands.}

\altaffiltext{11}{Subaru Telescope, 650 North A`ohoku Place, Hilo,
                  HI 96720, USA.}

\begin{abstract}

  We report monitoring observations of the T~Tauri star EX~Lupi
  during its outburst in 2008 in the CO fundamental band at
  4.6--5.0~$\mu$m. The observations were carried out at the {\it
    VLT} and the Subaru Telescope at six epochs from April to
  August 2008, covering the plateau of the outburst and the
  fading phase to a quiescent state. The line flux of CO
  emission declines with the visual brightness of the star and
  the continuum flux at 5~$\mu$m, but composed of two
  subcomponents that decay with different rates. The narrow line
  emission (50~km~s$^{-1}$ in FWHM) is near the systemic
  velocity of EX~Lupi.  These emission lines appear exclusively
  in v=1-0. The line widths translate to a characteristic
  orbiting radius of 0.4~AU. The broad line component (FWZI
  $\sim$ 150~km~s$^{-1}$) is highly excited upto v$\leq$6. The line
  flux of the component decreases faster than the narrow line
  emission. Simple modeling of the line profiles implies that
  the broad-line emitting gas is orbiting around the star at
  0.04--0.4~AU. The excitation state, the decay speed of the
  line flux, and the line profile, indicate that the broad-line
  emission component is physically distinct from the narrow-line
  emission component, and more tightly related to the outburst
  event.

\end{abstract}

\keywords{circumstellar matter --- planetary systems:
  protoplanetary disks --- stars: activity: other --- stars:
  pre-main sequence --- stars: formation --- stars: individual
  (EX Lupi)}

\section{Introduction}

EX~Lupi, the prototype of the class of the eruptive pre-main
sequence stars called EXors,\footnote{The group of low-mass
  stars with large variability is coined as EXors by
  \citet{her89} for its nature of the outburst similar to FU~Ori
  variables that are collectively called FUors.} underwent its 
largest outburst recorded in the past 50 years in early 2008
\citep[history of the observations of EX~Lupi and
EXors:][]{her07,her08}. The observational definition of EXor
variables relies strongly on the prototype EX~Lupi, which
spectroscopically looks like a classical T~Tauri star in the
quiescent phase \citep{her50,her01}, but flares by 1--3
magnitudes in the visible roughly once a decade, with each
outbreak lasting about a year \citep{her07}. The spectroscopic
behavior in the outburst is complicated, and sometimes related
to an expanding stellar shell \citep[e.g.][]{her89}. It is
increasingly accepted, however, that the circumstellar disk
plays the primary role in the outbursts, in which mass accreted
from the envelope is stalled somewhere in the disk, until it
reaches a critical surface density and triggers a disk
instability. 

\begin{figure*}
\includegraphics[width=0.8\textheight,angle=0]{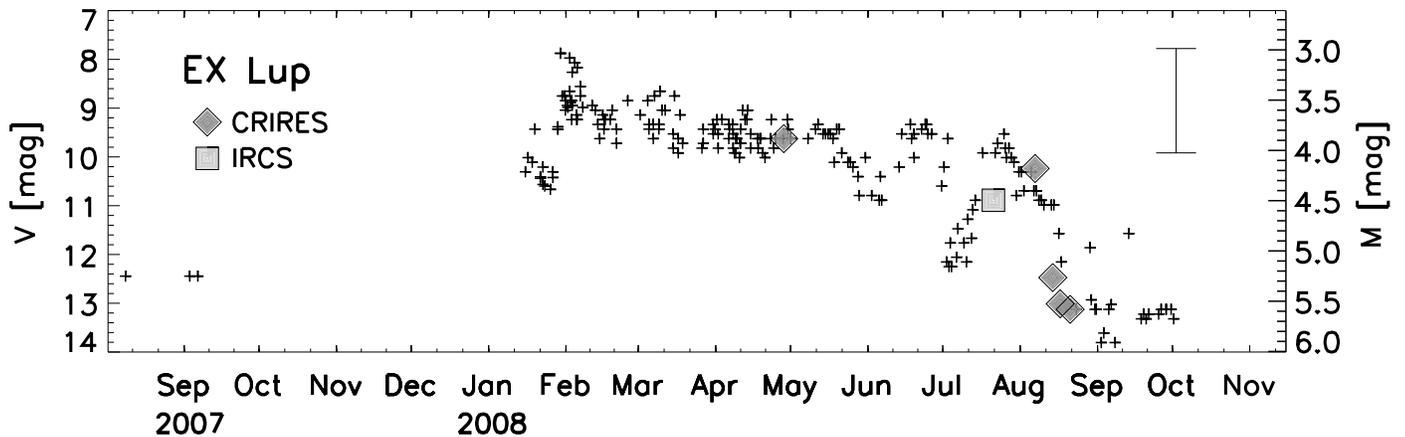}
\caption{Light curve in the visible (crosses; from AAVSO
  http://www.aavso.org/.) with the ordinate to the left, and in
  $M$ band (diamonds for CRIRES, and a square for IRCS) to the
  right. The axis of $M$ band magnitude is arbitrarily scaled to
  that of $V$ band. The $M$ band photometry is performed with
  the spectroscopic data, with the continuum level of the
  spectroscopic standard star as the photometric reference. The
  uncertainty of $M$ band photometry is $\pm$0.5~mag (the error
  bar shown in the upper right corner). \label{lc} }
\end{figure*}

The idea that EXor outbursts arise  inside the
accretion disk originates in their observational similarity 
with FUors. The most compelling evidence that the FUor outburst
has its origin in a disk is the split absorption lines in the
visible to the near-infrared with wavelength separation of two
absorption peaks gradually decreasing with the wavelength
\citep{har04}. This is most naturally accounted for when the
absorption lines at long wavelengths arise from the cool outer
part of the disk, where the orbital velocity is slower
\citep[e.g.][]{zhu09}. FUors and EXors are historically
considered as different classes of eruptive pre-main sequence
stars. However, both classes may 
represent a similar eruptive event, and differ only
quantitatively in terms of flare amplitude, outburst duration,
and outburst frequency. Recent outbursts of V1647~Ori
\citep[the central source of ``McNeal's Nebula'', e.g.][]{asp08} are 
recurrent like EXors, yet the amplitude of the outburst is more 
compatible with FUors, which fills the gap between these two
classes. The mass accreted in a single outburst and the
frequency of the outbursts indicates that an EXor/FUor star
acquires a substantial fraction of its mass solely during the
outbursts. It lends circumstantial support to the theory that
EXors/FUors are not a group of special variable stars, but a
state of the accretion phase that stars commonly experience during
their evolution \citep[e.g.][]{har96}. However, there are indeed
qualitative differences between FUors and normal T~Tauri stars,
such as the clustering nature, as FUors tend to be in isolation,
while other T~Tauri stars are in
clusters \citep{her03}. Moreover, the number of known FUors in the
Orion Nebula Cluster falls short if all class~I sources go
through the FUor phase with a finite duty cycle \citep{fed07},
although the number may increase with the inclusion of the
``FUor-like'' stars being identified recently \citep{gre08}. 
The current focus of EXor studies is to understand (1) what is
the mechanism of the disk instability, (2) what is the
relationship between EXors and FUors, and (3) what is the role
of EXor/FUor events in the context of star formation.

In this report, we will concentrate on the cause of the disk
instability and the physical scale of the outburst, as it often
gives a clue to the trigger of the instability. The most favored
model to date, thermal instability in a disk, is induced by
ionization, and subsequent opacity increase due to negative
hydrogen ions \citep{bel91,cla89}. The thermal instability is
therefore only operational within a limited disk radius, where
atomic hydrogen is subject to thermal ionization. If the
physical size of the outburst is larger, magnetohydrodynamic
(MHD) instability may be favored, as it predicts a massive
accretion in the magnetic dead zone at 1--10~AU
\citep{arm01}. If the outburst involves the whole disk, the
scenario of an unseen companions leading to global disk
instability may also work \citep{bon92}.

The vibrational transitions of CO are a unique probe of the
inner regions of disks (0.01--1~AU) due to their high critical
density ($n_{\rm cr}>10^{10}$~cm$^{-3}$) and high excitation
energy ($\Delta E_{\rm v=1-0}>3000$~K). Measurement of the
location of the emitting gas is relatively straightforward under
the assumption that the gas is in Keplerian rotation
\citep[e.g.][]{naj96}. We utilized multi-epoch observations of
CO vibrational transitions in EX Lupi to constrain where in the
disk the hot gas is located during the outburst, and how
  its physical property develops with time. The present study
is part of the first multi-wavelength campaign of EX~Lupi from
the optical \citep{sic10}, near-infrared \citep{kos10} to
mid-infrared wavelengths \citep{abr09, juh10} during the 2008
eruption. We will use $d=155$~pc as the distance to EX~Lupi from
\citet{lom08}, and the stellar parameters from \citet{gra05}
following \citet{sip09} (Table~\ref{t1}).

\section{Observations} 

The spectroscopic observations of the CO fundamental band were
performed at the {\it VLT} and the Subaru Telescope over a total
of 6 epochs (Figure~\ref{lc}). Half of the data were obtained in
the director's discretionary time (DDT) on 14, 17, and 21 August
2008 at the {\it VLT}. Two wavelength intervals of the CO
fundamental band (4.66--4.77~$\mu$m and 4.86--4.99~$\mu$m with
gaps between the four detector chips) were observed with CRIRES
\citep{kau04} to cover the low [$P$(1)--$P$(11)] and the high
      [$P$(21)--$P$(32)] rotational transitions of v=1-0 in
      order to provide good constraints on the rotational
      excitation temperature of the hot gas. The slit width was
      0\farcs2, corresponding to a nominal spectral resolution
      of $R$=100,000 ($\Delta v$=3~km~s$^{-1}$). The spectra
      were recorded by nodding the telescope along the slit to
      remove the sky background. The slit was oriented along
      P.A. = 0\degr, 180\degr, 90\degr~and 270\degr~by rotating
      the image rotator for the possible use of the data for
      spectroastrometry. Unfortunately, the achieved signal to
      noise ratio was not sufficient to extract any meaningful
      astrometric signal. The data obtained with different
      P.A. were simply summed and reduced as classical
      spectroscopic data. The total integration time was 12 and
      16 minute for each grating setting. The MACAO adaptive
      optics system \citep{bon04} was used with EX~Lupi itself
      as the guide star. Spectroscopic standard stars were
      observed at close airmass with a same instrument setup.

\begin{deluxetable}{lll}
  \tabletypesize{\normalsize} 

  \tablecaption{Stellar Parameters for EX~Lupi used in this paper.\label{t1}}
  \tablewidth{0pt}
\tablecolumns{3}
\tablehead{
\colhead{}&
\colhead{}&
\colhead{Reference}
}

\startdata
$M_\ast$ \dots & 0.6~$M_\odot$ &1.\\
$R_\ast$ \dots & 1.6~$R_\odot$ & 1. \\
$L_\ast$ \dots & 0.5~$L_\odot$ & 1. \\
$d$ \tablenotemark{a}\dots & 155~pc & 2. \\
$v_{\rm LSR}$\tablenotemark{b} \dots & $\approx$0~km~s$^{-1}$ & 3. \\
\enddata

\tablenotetext{a}{Distance to EX~Lupi.}
\tablenotetext{b}{Systemic velocity as measured with optical
  spectroscopy.}
\tablecomments{Reference: 1. \citet{gra05}, 2. \citet{lom08}, 3. \citet{her03}.}

\end{deluxetable}

EX~Lupi was also observed by CRIRES at the {\it VLT} in the
framework of the ESO large program 179.C-0151 on April 28 and
August 7 in 2008. The observations covered the CO spectrum at
4.65--4.77~$\mu$m with additional coverage at 4.81--4.90~$\mu$m
in the August observations. The slit width and the AO settings
were the same as during the DDT observations. The total exposure
times were about 20 min per grating setting in both of the
nights.

IRCS \citep{tok98} at the Subaru Telescope was used to obtain CO
fundamental band spectra of EX~Lupi on 21 July 2008. The angle
of the echelle grating was set so that the spectral interval of
4.65--4.74~$\mu$m was covered by the detector. The 0\farcs15
slit was used resulting in a spectral resolution of $R$=20,000
($\Delta v=15$km~s$^{-1}$). The spectra were recorded by nodding
the tip-tilt mirror inside the AO system 3\farcs4 along the slit
to remove the sky background. The observation was performed
through a large airmass ($\sec{z} \sim$ 2) with 5 times
lower spectral resolution than CRIRES. Thus, the data quality is
significantly lower. The total integration time was 14
minutes. The adaptive optics system \citep{hay08} was also used
here with EX~Lupi as the guide star. The observations of all 6
epochs are summarized in Table~\ref{t2}.

One-dimensional spectra were extracted from the CRIRES dataset
using the crires\_spec\_jitter recipe\footnote{CRIRES Pipeline
  User Manual VLT-MAN-ESO-19500-4406.} on the ESO gasgano
platform\footnote{http://www.eso.org/sci/data-processing/software/gasgano/.}. A
custom-written IDL code was used to divide the spectra of
EX~Lupi by standard star spectra to eliminate telluric
absorption lines. In the cases where the precipitable water
vapor varied during the night, the water absorption was removed
from both the object and the standard spectra beforehand by
dividing them by the ATRAN atmospheric transmission model
\citep{lor92}. The IDL code further corrected wavelength offsets
between EX~Lupi and the spectroscopic standard spectra, slight
differences in spectral resolution, and mismatches of the
airmass. The wavelength calibration was performed by maximizing
the cross-correlation between the observed spectra and the
atmospheric transmission curve. The wavelength calibration error
depends on the density of the telluric lines, and is less than a
few pixels ($\sim$5~km~s$^{-1}$).

The IRCS data were reduced in the same way, except that the
IRAF\footnote{IRAF is distributed by the National Optical
  Astronomy Observatories, which are operated by the Association
  of Universities for Research in Astronomy, Inc., under
  cooperative agreement with the National Science Foundation.}
aperture extraction package was used to extract the
one-dimensional spectra. 
 
EX~Lupi is an M0 star in its quiescent phase with CO bandhead
absorption at 2.3~$\mu$m \citep{sip09}.  Nevertheless, the
photospheric correction for the CO absorption lines was not
performed at 4.7~$\mu$m, as the photospheric temperature during
the outburst is unknown. The contribution of the stellar
photosphere at 5~$\mu$m is less than 20\% in the quiescent phase
\citep{sip09}, and expected to be less during the outburst
\citep{juh10}.

\setlength{\tabcolsep}{0.015in}
\begin{deluxetable*}{llcccccccccc}
\tabletypesize{\scriptsize}
\tablecolumns{12}
\tablewidth{0pt}
\tablecaption{Journal of observation.\label{t2}}

\tablehead{
\colhead{Epoch}&
\colhead{Program}&
\colhead{JD}&
\colhead{{\it V}}&
\colhead{{\it M}}&
\colhead{Spectrograph}&
\colhead{Telescope}&
\colhead{Velocity}&
\colhead{Wavelength}&
\colhead{Integration}&
\colhead{Standard}&
\colhead{Spectral}\\
\colhead{}&
\colhead{}&
\colhead{}&
\colhead{[mag]\tablenotemark{a}}&
\colhead{[mag]\tablenotemark{b}}&
\colhead{}&
\colhead{}&
\colhead{Resolution}&
\colhead{Coverage}&
\colhead{Time [min]\tablenotemark{c}}&
\colhead{Star}&
\colhead{Type}}

\startdata
28 Apr 2008 & 179.C-0151(D) &2454585
&  9.7  &    3.88
&  CRIRES & VLT 
&3~km~s$^{-1}$  & 4.65--4.77~$\mu$m & 20 &HR~6084 & B1~III\\

21 Jul 2008 & DDT           &2454669 
& 10.0 &  4.50  
& IRCS  & Subaru&15~km~s$^{-1}$ & 4.65--4.74~$\mu$m & 14
& HR~5987 & B2.5Vn\\

~7 Aug 2008  & 179.C-0151(D) &2454686 
& 10.8 & 4.18
& CRIRES & VLT
&3~km~s$^{-1}$  & 4.65-4.77$\mu$m, 4.80-4.90$\mu$m & 20
& HR~6084 & B1~III \\  

14 Aug 2008  & 281.C-4031(D) &2454693 
& 11.1 & 5.27
& CRIRES & VLT
&3~km~s$^{-1}$  & 4.66--4.77$\mu$m\tablenotemark{d}
4.86-4.99$\mu$m\tablenotemark{d}  & 12, 16 & HR~6508& B2IV\\ 

17 Aug 2008  & 281.C-4031(E) &2454696 
& 12.3 & 5.53
& CRIRES & VLT
&3~km~s$^{-1}$  & 4.66-4.77$\mu$m\tablenotemark{d} 
4.86-4.99$\mu$m\tablenotemark{d}  & 12, 16 & HR~6508& B2IV\\ 

21 Aug 2008 & 281.C-4031(F) &2454700 
& 13.3 & 5.58
& CRIRES & VLT
&3~km~s$^{-1}$  & 4.66-4.77$\mu$m\tablenotemark{d} 
4.86-4.99$\mu$m\tablenotemark{d} & 12, 16 & HR~6508& B2IV\\ 
\enddata

\tablenotetext{a}{From AAVSO http://www.aavso.org/.}
\tablenotetext{b}{Photometry is done  with respect to the
  spectroscopic standard stars. The accuracy is $\pm$~0.5~mag.}
\tablenotetext{c}{For one grating setting.}
\tablenotetext{d}{With gaps between the four detector chips.}
\end{deluxetable*}

Since there is no simultaneous photometry available at 5~$\mu$m,
the absolute flux calibration was performed against the
spectroscopic standard stars HR~6084, HR~5987, and HR~6508.
HR~6084 is a B1~III star with an $M$ band magnitude of
2.42~mag\footnote{http://www.jach.hawaii.edu/UKIRT/astronomy/calib/phot\_ cal/bright\_stds.html}. HR~5987
and HR~6508 are B2.5~Vn and B2~IV stars with $K$ magnitudes of
4.70 and 3.18~mag, respectively, but without accurate $M$ band
brightness documented. We applied a color correction of
$K-M=-0.05$~mag taken from $K-L$ of early B stars [\citet{tok99}
with the original from \citet{koo83}], that gives $M=4.75$ and
3.23~mag, respectively. The precise correction is unimportant,
since the overall photometric accuracy is only $\pm$0.5~mag. The
main source of photometric error is the variable slit
transmission at the time of the observation up to a factor of
three even in a pair of consecutive frames. The variable
atmospheric transmission during the nights adds another 30\% to
the photometric uncertainty. EX~Lupi becomes fainter from 3.9 to
5.6~mag in the $M$ band over 4 months (Table~\ref{t2}). The
photometric value on 28 April is surprisingly bright, as Spitzer
spectroscopy in April 2008 indicates 1.5 Jy at 5~$\mu$m, or
$M\approx$5.0~mag \citep{juh10}. Since then, the star became
fainter at optical wavelengths\footnote{AAVSO
  http://www.aavso.org/.}, albeit with significant variability
on timescales of days to weeks. The data counts in the raw
frames are at least consistent with those of $M$=3.9--5.6 mag
stars, as simulated by the exposure time calculator of
CRIRES\footnote{http://www.eso.org/observing/etc/.}.

\section{Analysis}
\subsection{Two Emission Components}

\begin{figure*}
\begin{center}
\includegraphics[width=0.64\textheight,angle=0]{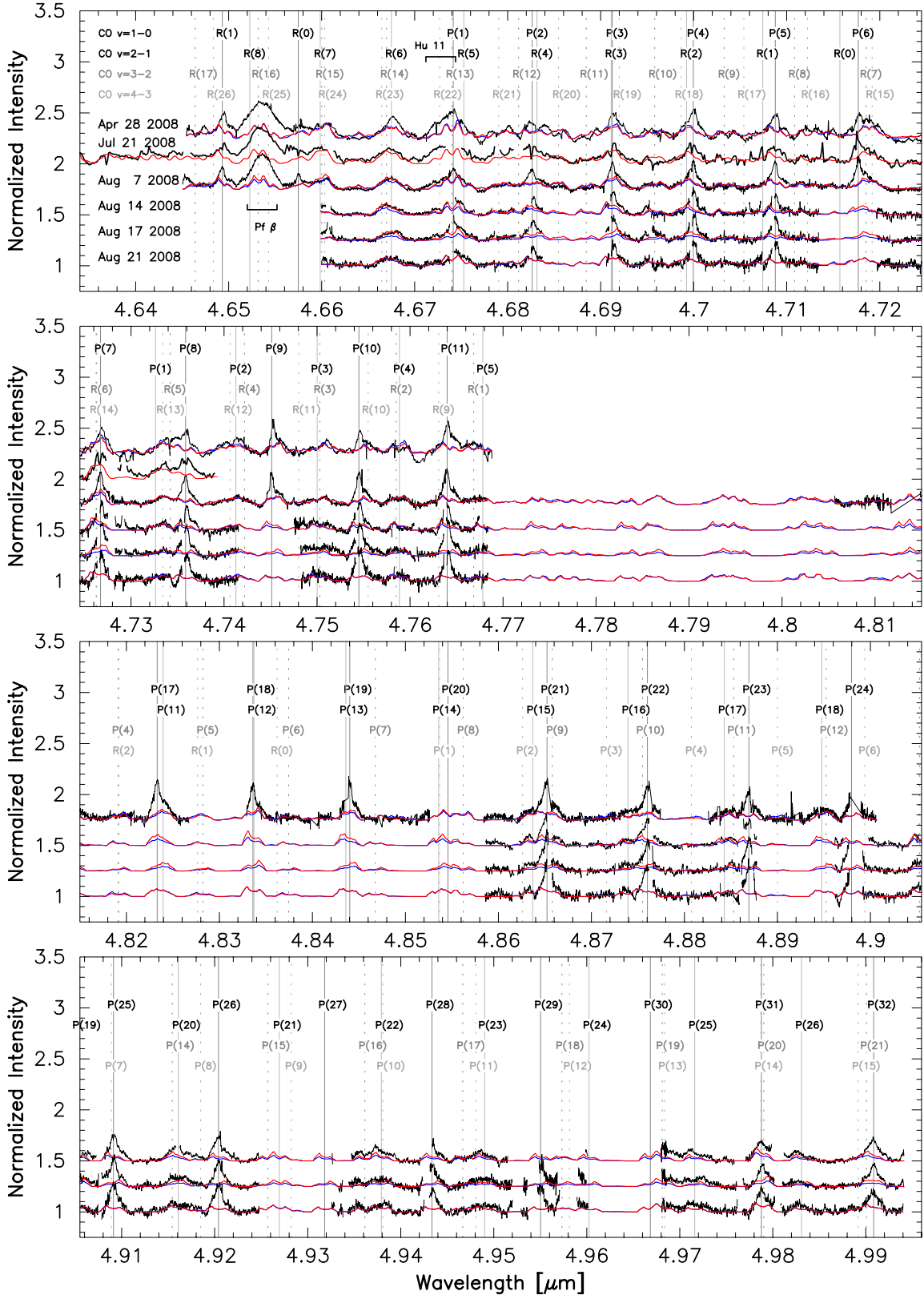}
\end{center}
\caption{Spectra of EX~Lupi at the CO fundamental band over six
  epochs from 28 April to 21 August 2008. All spectra were
  obtained by CRIRES at the {\it VLT} ($R$=100,000) except on 21
  July, when IRCS at Subaru Telescope was used with 5 times
  lower spectral resolution ($R$=20,000). The spectra are
  normalized to the continuum level. Two emission components are
  immediately noticeable: one broad emission that fades away in
  August, and narrow line emission. Two model spectra for the
  outburst component are shown overlaid. The blue models are
  calculated with the excitation temperatures that fit the
  population diagrams in Fig.~\ref{pp} and the constant column
  densities $2\times10^{20}$~cm$^{-2}$.  The red models are
  calculated by fixing the excitation temperature at $T_{\rm
    x}=4500$~K, which is the dissociation limit of the
  molecules. The two models supposedly represent the extreme
  cases meant to illustrate how the thermodynamical parameters
  are degenerated. The model parameters are listed in
  Table~\ref{t3}. The spectra are not covered continuously
  because of the gaps between the detector chips in
  CRIRES. \label{sp1} }
\end{figure*}

\begin{figure}
\begin{center}
\includegraphics[height=0.45\textheight,angle=0]{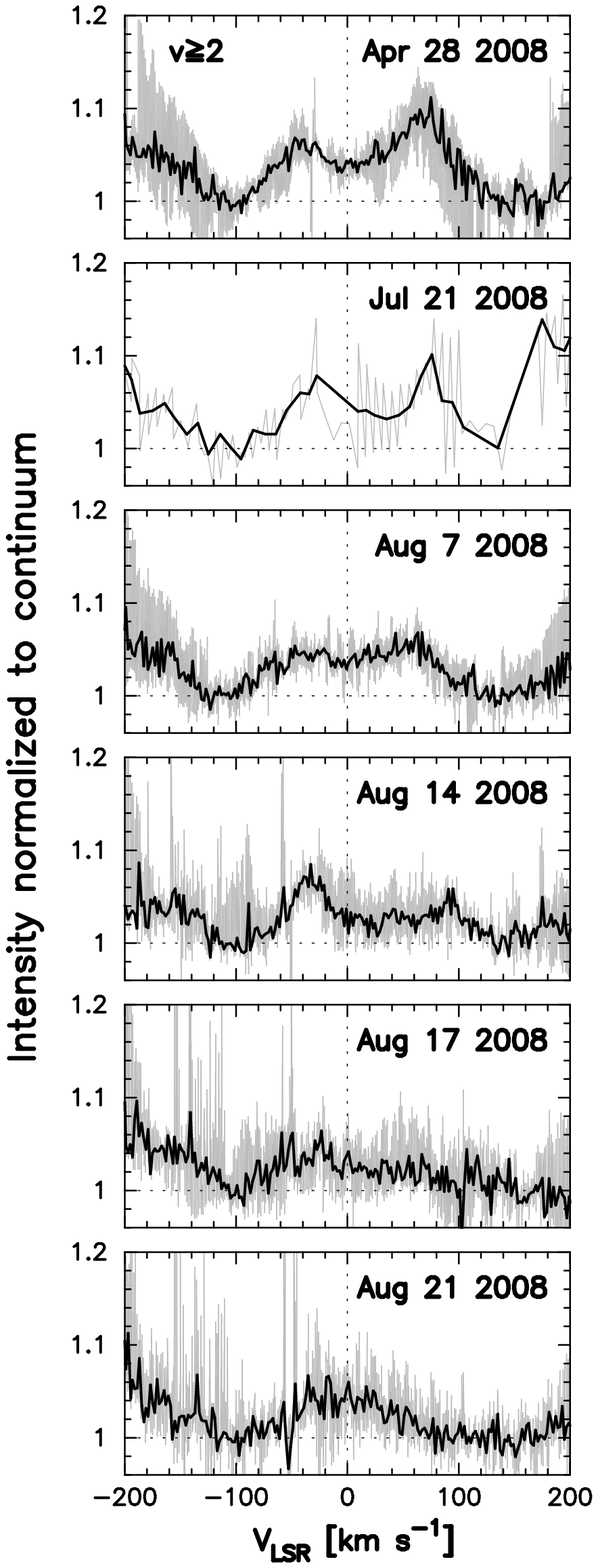}
\includegraphics[height=0.45\textheight,angle=0]{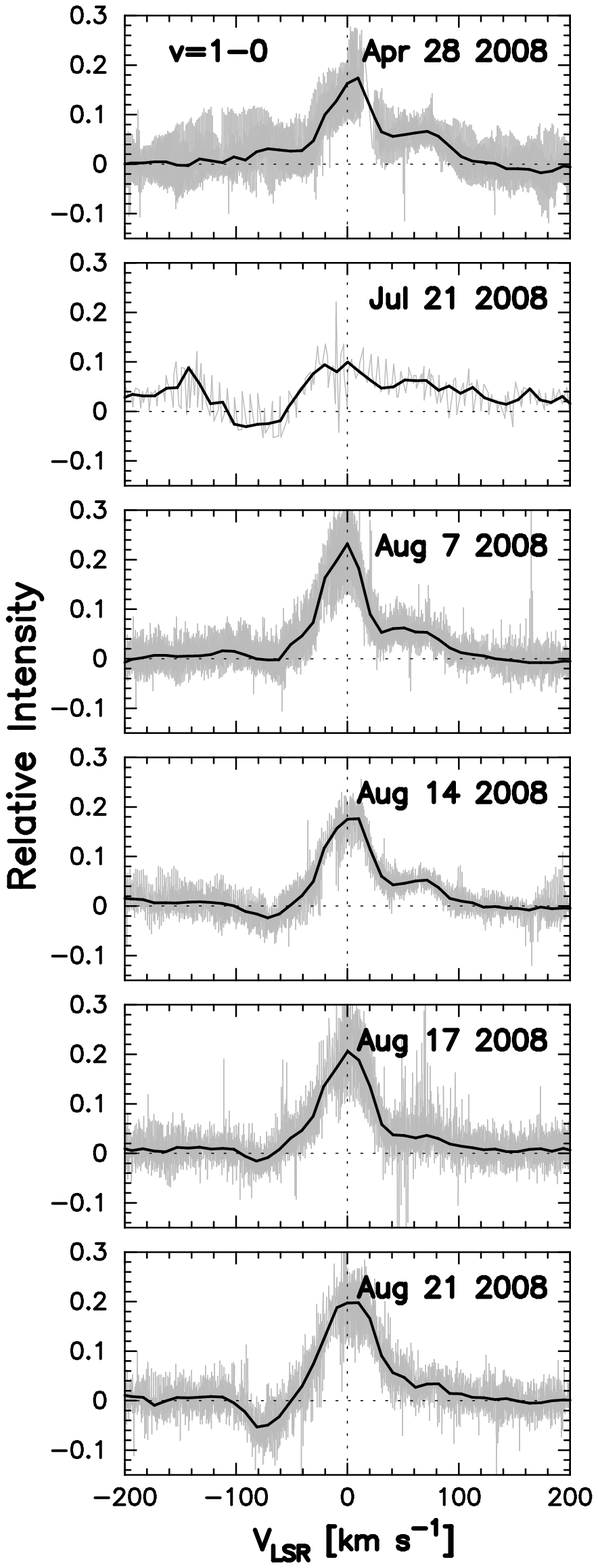}
\end{center}
\caption{Left: composite line profiles of the vibrationally
  excited transitions (v$\ge$2). The gray lines are the overlay
  of CO $P$(3) v=2-1 4.750~$\mu$m, CO $P$(4) v=2-1 4.759~$\mu$m,
  $R$(9) v=3-2 4.703~$\mu$m, and $R$(21) v=4-3 4.679~$\mu$m. The
  black line is the averaged spectrum. The line emission is
  double-peaked. Right: composite line profiles of CO v=1-0
  after the outburst models of $T_{\rm x}=4500$~K in
  Figure~\ref{sp1} are subtracted. The gray lines are overlay of
  v=1-0 lines within the spectral coverage, and the black lines
  are the averaged spectrum of them. A blue depression shows up
  at the velocity of $-$80~km~s$^{-1}$ in August, and becomes
  deeper later.  We attribute it to the disk wind started at the
  late phase of the outburst. \label{dp}}
\end{figure}

All spectra from the 6 epochs are shown in Fig.~\ref{sp1} after
having been normalized to the continuum and corrected to the
local standard of the rest. Two spectral components are
immediately noticeable in the CO spectra, one broad [full width
at zero intensity (FWZI) $\sim 150$~km~s$^{-1}$] and one narrow
[full width at half maximum (FWHM)$\sim 50$~km~s$^{-1}$]
overlaid on each other. The broad line emission is
  characterized by the hot vibrational transitions up to v=6
  confirmed.  This is in agreement with the first overtone band
  in emission seen at least up to v=5-3 in the SINFONI spectra
  \citep{kos10} observed on 25-31 July prior to the CRIRES DDT
  runs during the outburst.  In order to illustrate the broad
line profiles of the former, a few lines from v$\ge$2 levels,
which are relatively isolated from others, such as $P$(3) v=2-1
at 4.750~$\mu$m and $R$(9) v=3-2 at 4.703~$\mu$m, are combined
and shown in the left panel of Fig.~\ref{dp}. The broad emission
lines are most clearly seen in the April 28 data, taken during
the outburst, a few months after the optical maximum.  

The sharp emission lines on top of the broad component are close
to the systemic velocity of EX~Lupi \citep[$v_{\rm LSR}\approx
0$~km~s$^{-1}$; ][]{her03}. The intensity of the narrow lines is
10--20\% above the continuum level throughout the
observations. The line profile of the emission lines does not
change significantly during and at the end of the outburst
(Fig.~\ref{dp}, right).  The line widths, markedly smaller than the
broad line component, imply that the emission arises from
regions of the disk outside where the broad and highly-excited
emission lines comes from.

The spectral profile, composed of two distinct kinematic
components, is somewhat similar to \ion{Fe}{2} at
5018~\AA~observed with the HIRES spectrograph ($R=48,000$) at
Keck in 1998 when EX~Lupi was at the maximum in a previous
outburst. \citet{her07} attributed the broad line emission to
the accretion funnels, as the \ion{Fe}{2} emission had inverse
P~Cyg profile with broad absorption longward at
$+300$~km~s$^{-1}$. The velocity widths of the broad and the
narrow components are 140~km~s$^{-1}$ and 23~km~s$^{-1}$ in
\ion{Fe}{2} 5018~\AA, similar to the CO lines at 4.7~$\mu$m. We
looked for similar systematic redward absorption in the CO
lines, and found that they are not present in either epoch of
the observations. Moreover, as we discuss later, the line
profile in the outburst is double-peaked, which implies that the
emitting gas is orbiting around the star. We conclude that the
origin of the broad CO emission is not the stellar photosphere
or accretion columns, but the circumstellar disk. We will
deconvolve the broad and the narrow emission components using a
simple slab model in the next sections.

\subsection{Population Diagrams}

\begin{figure}
\begin{center}
\includegraphics[height=0.8\textheight,angle=0]{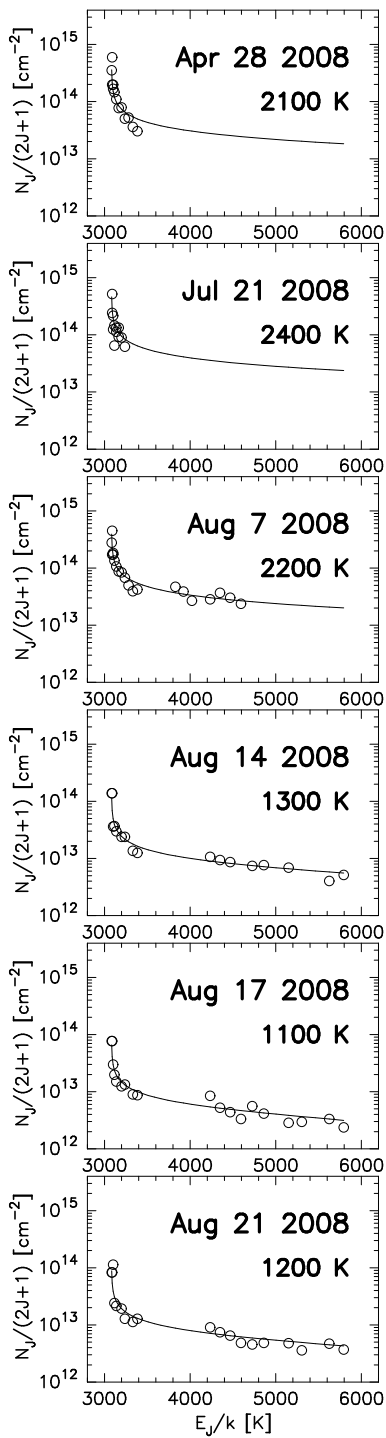}
\caption{Population diagram of CO v=1-0.  The line flux is
  calculated by integrating over $\pm$120~km~s$^{-1}$ of the
  emission lines. The emission is optically thick as is shown by
  the turnovers at $E_J/k=3000$--$3400$~K. The thermodynamical
 models with assumed line profile (Gaussian profile with
  $\sigma_{v}=5$~km~s$^{-1}$) are shown overlaid. 
  \label{pp}}
\end{center}
\end{figure}

First, population diagrams were constructed to obtain an
overview of the physical property of the hot emitting gas
in the disk. Equivalent widths of the emission lines were
measured integrating v=1-0 $P$-branch lines over
$\pm$120~km~s$^{-1}$.
They thus include v=1-0 emission of both broad and narrow
components without correcting for the overlapping higher
vibrational transitions. The column density at each rotational
level is calculated as $h\nu\,N_J A_\nu \cdot \pi R_s^2 = 4\pi
d^2\, W_\nu f_0$ from the equivalent widths $W_\nu$. $f_0$ is
the continuum flux coming from the $M$-band photometry. The
specific radius of the emitting area $R_s$ is set to 0.2~AU, so
that it is roughly consistent with the line width of the broad
line emission. The column density normalized by the degeneracy
of the levels $N_J/(2J+1)$ is plotted against the energy levels
$E_{J}/k$ in Fig.~\ref{pp}. The observed population diagrams
show turnovers at $E_{J}/k = 3200$--$3400$~K, which testifies
the emitting gas is optically thick. The vertical offset of the
diagrams over 6 epochs spanned roughly an order of magnitude.
The population diagram was initially fit by a thermodynamical
model with $T_{\rm x}$ and $N_{\rm CO}$ as free parameters.
However, in order to reproduce the sequences of $N_J$ in the
optical thick regime with an order of magnitude dispersion,
$N_{\rm CO}$ has to be adjusted by several orders of magnitude
up to $10^{25}$~cm$^{-2}$, which is apparently overstretching
the capability of the model. We therefore fixed the column
density to the minimum value of 6 epochs ($N_{\rm CO}=
2\times10^{20}$~cm$^{-2}$) found in the first path of the
fitting, and tried to match the absolute levels of $N_J$ by
changing the size of the emitting area. The column density
$2\times10^{20}$~cm$^{-2}$ is close to the upper limit of the
physically feasible value of the relatively small area of the
disk ($R_s\sim 0.1$~AU, $H/R$=0.1, [C/H]=10$^{-4}$, $n_{\rm
  H}=10^{12}$~cm$^{-3}$).
The fractional column density at v=1 level was calculated with
the vibrational temperature $T_{\rm v}$ set equal to $T_{\rm
  x}$. The best-fit $T_{\rm x}$ is constrained within
1100--2400~K and tabulated in Table~\ref{t3} with the
corresponding emitting radii of the disk $R_s$. The line profile
is assumed to be a Gaussian function of
$\sigma_v=5$~km~s$^{-1}$.

\subsection{Slab Model}
Second, we used a simple slab model to incorporate the geometry
and the rotation of the disk to reproduce the observed
spectra. The slab model is a gas disk with no geometrical
thickness, and meant to reproduce the underlying broad line
emission only. The fundamental transition of CO is used as the
sole opacity source at the wavelength concerned. No isotopomers
are included, as this does not change the quality of the fit
within the given signal-to-noise ratio of the
spectra. Rotational levels up to $J= 100$, and vibrational
levels up to v=8 are taken into account. The wavelengths, energy
levels, and Einstein $A_\nu$ coefficients are taken from
\citet{goo94}. Ro-vibrational levels are populated according to
the single-temperature Boltzmann distribution.
Thermal broadening at the rotational temperature
is taken as the only source of the intrinsic line width. The
input parameters of thermodynamics, the total
column density $N_{\rm CO}$, and the rotational and vibrational
temperatures $T_{\rm x}$ and $T_{\rm v}$, 
are kept constant without radial dependency. The resulting
spectrum is calculated as $I_\nu = \frac{1}{4\pi}
S_\nu(1-e^{-\tau_\nu})$, where $S_\nu$ is the line source
function $A_{21} n_2 / (B_{12} n_{1} - B_{21} n_{2})$, and
$\tau_\nu = \frac{h\nu_{12}}{4 \pi} (n_{1} B_{12} - n_{2}
B_{21}) \phi_\nu / \cos{i} $ is the opacity with $\phi_\nu$
being the dimensionless line profile, and $i$ being the disk
inclination angle.

We tested a model with the opacity of the dust grains uniformly
mixed with the gas. If we take the mass absorption coefficient
$\kappa (\lambda={\rm 5 ~\mu m}) = $2000~cm$^2$~g$^{-1}$, as
calculated by \citet{oss94}, the line emission of CO will be
entirely suppressed by the dust continuum emission, unless the
gas to dust mass ratio is larger than 10,000. We infer that the
continuum emission at 5~$\mu$m comes from the region outside the
broad line emission region.  The continuum emission is therefore
removed from the observed spectra, and only the remaining line
emission part is produced by the slab model, without dust
continuum opacity in the radiation.

The disk is divided into annuli from the inner ($R_i$) to the
outer disk ($R_o$) boundary in the way that $\log{r_{j+1}} -
\log{r_{j}}$ is constant. The ro-vibrational emission spectrum
is calculated in a small segment of each annulus with the given
rotational and vibrational temperatures. The line spectra are
then shifted by the radial velocity of the Keplerian rotation at
the given azimuthal location, and added together. The stellar
mass is assumed to be $M_\ast = $0.6$M_\odot$ \citep{gra05}. The
line emission spectra at the different radii are combined with
an arbitrary weighting function $I(r)
=I(R_i)(\frac{r}{R_i})^{-\alpha}$, where $\alpha$ is fixed to
2.5 in the present modeling. The inclination angles between
40\degr\,\,and 50\degr\,\,give reasonably good fits to the
spectra at all the epochs. We therefore fixed it to
$i=$45\degr\,\,without further adjustment. The disk model of
\citet{sip09}, which reproduces the optical-infrared SED of
EX~Lupi in the quiescent phase favors face-on geometry with the
most plausible inclination angle being
$i=$20\degr. \citet{sip09} found, however, that changing the
inclination angle from 0\degr\, to 40\degr\, changes the output
SED little; as the SED is relatively insensitive to the
inclination angle, unless the disk is heavily optically
thick. The inclination angle we adopt here is near the upper
limit of this range.

\setlength{\tabcolsep}{4pt}
\begin{deluxetable*}{lccc|ccccccccc}
\tabletypesize{\scriptsize}
\tablecolumns{13}
\tablewidth{0pt}
\tablecaption{Disk parameters from our modeling.\label{t3}}

\tablehead{
\colhead{}&
\colhead{$R_s^\prime$\tablenotemark{a}} &
\colhead{$N_{\rm CO}^\prime$\tablenotemark{b}} &
\colhead{$T_{\rm x}^\prime$\tablenotemark{c}} &
\colhead{$i$\tablenotemark{d}}&
\colhead{$R_i$\tablenotemark{e}}&
\colhead{$R_o$\tablenotemark{f}}&
\colhead{$\alpha$\tablenotemark{g}}&
\colhead{$N_{\rm CO}$\tablenotemark{h}} &
\colhead{$T_{\rm v}$\tablenotemark{i}}&
\colhead{$T_{\rm x}$\tablenotemark{j}}&
\colhead{$a$\tablenotemark{k}}&
\colhead{$F_\lambda$\tablenotemark{l}} \\
\colhead{Epoch}&
\colhead{[AU]}&
\colhead{[$\times$10$^{19}$cm$^{-2}$]}&
\colhead{[K]}&
\colhead{[\degr]}&
\colhead{[AU]}&
\colhead{[AU]}&
\colhead{}&
\colhead{[$\times$10$^{19}$cm$^{-2}$]}&
\colhead{[K]}&
\colhead{[K]}&
\colhead{}&
\colhead{[ $\times$10$^{-13}$ W m$^{-2}\mu$m$^{-1}$]}}

\startdata
 28 Apr 2008  & 0.5&20 &2100 & 45 & 0.05 & 0.3     & 2.5 & 6 & 3000 & 4500 & 0.7 & 5.9 \\
 21 Jul 2008  & 0.3 &20 & 2400 & 45 & 0.05 & 0.3      & 2.5 & 4 & 2500 & 4500 & 1.0 & 3.4 \\
 ~7 Aug 2008  & 0.4 &20 & 2200 & 45 & 0.055  & 0.4 & 2.5 & 2 & 2700 & 4500 & 0.8 & 4.5 \\
 14 Aug 2008  & 0.4 &20 & 1300 & 45 & 0.05  & 0.3   & 2.5 & 1.5 & 2000 & 4500 & $-$0.6 & 1.7\\
 17 Aug 2008   & 0.5 &20 & 1100 & 45 & 0.04 & 0.3   & 2.5 & 4 & 1800 & 4500 & 0.6 & 1.3 \\
 21 Aug 2008   & 0.4 &20 & 1200 & 45 & 0.04 & 0.2   & 2.5 & 2 & 2000 & 4500 & $-$0.8 & 1.2 \\
\enddata

\tablenotetext{a}{Radius of the emitting area from the
  population diagram in Fig.~\ref{pp}.}  \tablenotetext{b}{Total
  column density of CO fixed to $2\times20$~cm$^{-2}$.}
\tablenotetext{b}{Rotational excitation temperature from the
  population diagram in Fig.~\ref{pp}.}

\tablenotetext{c}{Inclination angle from face-on.}
\tablenotetext{d}{Inner radius.}
\tablenotetext{e}{Outer radius.}
\tablenotetext{f}{Weight function of the intensity at different radius
$I(r)=I(R_i)(\frac{r}{R_i})^{-\alpha}$.}
\tablenotetext{g}{Total column density of CO.}
\tablenotetext{h}{Vibrational excitation temperature.}
\tablenotetext{i}{Rotational excitation temperature, fixed to 4500~K.}
\tablenotetext{j}{Asymmetry factor of line profile.}
\tablenotetext{k}{Flux density at the continuum level.}

\end{deluxetable*}

\section{Discussion}

\subsection{Spectral Components}

\subsubsection{Broad Line Emission} 

The disk parameters of the broad line emission that we modeled
in the previous section, $i$, $R_i$, $N_{\rm CO}$, $T_{\rm x}$
and $T_{\rm v}$ are not independent, but interlocked in a
non-trivial way. The line width of $v \sin{i} \sim
150$~km~s$^{-1}$ on Apr 28 testifies that the radius of the
inner edge of the disk is not larger than 0.09~AU. On the other
hand, the disk cannot be too small in order to account for the
absolute line flux. In the cases when the line profile is not
simple and the signal-to-noise ratio is not high, such as the
present case, the photometric calibration is important because
the absolute line flux effectively constrains the size of the
emitting region and therefore the disk radius. If we take the
photometry at 5~$\mu$m derived based on the spectroscopic data
as face value, the single line flux ranges from 10$^{-14}$ to
10$^{-13}$~erg~cm$^{-2}$. The maximum flux density that we could
retrieve from a disk at $d$ pc away is $f_\nu \approx \pi r^2
B_\nu(T_{\rm b})/d^2$, where $B_\nu$ is the Planck
function. Given the blackbody temperature of $T_{\rm b}$ =
2000~K, the distance of $d =155$~pc, and the line width of
100~km~s$^{-1}$, we will get $f_\nu <
10^{-14}$~erg~s$^{-1}$cm$^{-2}$, if we assume a disk radius of
0.01~AU. This is too small to account for the observed line
flux, therefore the radius of the inner hot disk cannot be
smaller. The lower limit for the disk radius then calls for the
inclination angle to be larger than $i >$30\degr~to match the
observed line width.

We first applied $T_{\rm x}$ and $N_{\rm CO}$ to the models as
determined by the optically thick population diagrams, and
optimized gas radius to fit the line profiles best. The
vibrational temperatures $T_{\rm v}$ are set equal to $T_{\rm
  x}$. The results underestimate the absolute line flux, because
the outer radius $R_o$ that is consistent with the kinematics in
the line profiles is larger than the specific emitting radius
$R_s$ we used to reproduce the population diagrams. The model
spectra are normalized by the square of $R_o/R_s$ and shown in
Fig.~\ref{sp1} in blue lines.
The absolute line flux is reproduced without such an arbitrary
scaling, by increasing the excitation temperature, allowing it
to differ from $T_{\rm v}$. However, when the emitting gas is
hotter than 3000~K, the line ratios of different rotational
transitions diminish in a short wavelength interval. It becomes
less straightforward to determine $T_{\rm x}$ uniquely.
In the optically thick regime, increasing $N_{\rm CO}$ and
$T_{\rm x}$ imposes similar effects on the output spectra. In
order to cope with the degeneracy, and avoid increasing the
optical depth any further, which our simple model cannot handle,
the rotational temperature was set to 4500~K, as the thermal
dissociation temperature of CO is 4500--4600~K
\citep{tat66}. $T_{\rm v}$ is adjusted so that the model
reproduce the absolute line flux under the condition of $T_{\rm
  x}=4500$~K. The second model with $T_{\rm x}=$4500~K is shown
in Fig.~\ref{sp1} in red lines. The two solutions
(Table~\ref{t3}) can be regarded as the two extreme cases of the
spectral models, and provide a fair idea how the model is
affected by the degeneracy of the parameters. The best-fit line
profile of a disk in Keplerian rotation sets the inner and the
outer radius at 0.04--0.05~AU ($=$5--7~$R_\ast$) and
0.2--0.4~AU, respectively, for all epochs with the fixed
inclination angle of $i=$45\degr.  These geometric parameters
are commonly used in the two representative models discussed
above, and less subjected to the degeneracy.  The optical depth
is $\tau_\nu=$0--1000 at the line center in the model with
$T_{\rm x}=T_{\rm v}$, and $\tau_\nu=$0--100 for $T_{\rm
  x}=4500$~K, assuming the line is only thermal broadened.

Although the heavy overlap of the lines makes it difficult to
discern the shape clearly, the line profile looks double-peaked
(Fig.~\ref{dp}). This is the first time that the line profile is
found to be double peaked in EXors.  It lends support to the
underlying similarity of the two classes of variable stars,
FUors and EXors, and the origin of EXor outbursts associated
with the circumstellar disk. Hartmann and his school
\citep[e.g.][]{har96,har04,zhu09} attribute the split absorption
lines of FUors, unusually broad for a stellar photosphere, to
the gas in Keplerian rotation in unstable disks. The interval of
line splitting is smaller in the near-infrared than in the
visible. This is naturally explained if the absorption lines at
longer wavelengths arise from the cooler, more distant,
slow-rotating region \citep{zhu09}. On the other hand,
\citet{her03} and \citet{pet08} found no correlation between the
line widths and the excitation potentials of the visible
absorption lines of FU~Ori, which should be present if the lines
arise from the disk, and the disk has a temperature
gradient. They argued against the disk origin of the optical
lines, and claimed that the split lines are better reproduced by
a large cool spot at the polar region on a rapidly rotating
star. Moreover, the absorption lines of FUors are not perfectly
Keplerian, but have a rectangular shape at the bottom
\citep[e.g.][]{har04}. It is hard to distinguish whether the
unusual line shape represents the turbulence in the upper layer
of the disk through which the absorption takes place, or a
particular morphology and geometry of the cool spot on the
surface of a star \citep[e.g.][]{pet08}. The line profile of
EX~Lupi is found to be double peaked, but in emission. The
double peaked emission lines would not be compatible with the
stellar spot hypothesis, unless the stellar surface is extended
to at least a few times the stellar radius.

The line profiles are asymmetric between the blue and the red
peaks. Figure~\ref{as} shows the blow-up view of the spectra
near $R$(6) v=2-1 (4.6675~$\mu$m). There are minor contributions
from $R$(14) v=3-2 and $R$(23) v=4-3 at 4.6665~$\mu$m and
4.6670~$\mu$m, respectively. The asymmetric line profile,
however, cannot be fully explained by changing the vibrational
temperature, thus the relative contribution of the three
lines. The line profile is fitted better if we artificially
introduce intrinsic asymmetry in the Keplerian line profile by
modulating the azimuthal intensity distribution sinusoidally
[$I_\nu^\prime(\theta)=I_\nu (\theta)
(1+(1-a)\cdot\sin{\theta})$, where $a$ is the asymmetric
factor] by down to $a=$0.6. Asymmetries in infrared CO line
emission have been found by \citet{got06} in HD~141569~A, where
the v=2-1 emission is three times brighter in the northern
disk than in the southern disk in the spatially resolved
spectroscopy. \citet{pon08} proposed that the transition disks
around SR~21 and HD~135344B have azimuthal structures, as the
spectroastrometric signal of CO v=1-0 is asymmetric, which is
consistent with the millimeter dust continuum imaging
\citep{bro09}.  The line asymmetry of EX~Lupi varies with time.

\subsubsection{Narrow Line Emission}

The single peak profile of the narrow line emission with
50~km~s$^{-1}$ line-width in FWHM defies pure Keplerian rotation
modeling, if the line intensity decreases as a power law of the
radius. Instead, the line width is simply taken as the orbital
velocity of the gas, and is translated to the disk radius of
0.4~AU, using the same inclination angle as the broad line
emission. FWHM was used instead of FWZI, as the latter is hard
to measure unambiguously because of the overlap with the broad
line components. The disk radius above does not mean the inner
truncation, but a characteristic radius where the narrow
line emission arise from. The quiescent line emission comes
exclusively from the v=1-0 transition with no obvious higher
vibrational transitions. Except the emergence of absorption at
the blue shoulder implying a disk wind, no significant changes
are found either in the equivalent width or in the line shape
(Fig.~\ref{ew}; Fig.~\ref{dp}, right). Note that the small
equivalent width on July 21 is of less significance than other
epochs, as the data suffer from the low spectral resolution and
the limited signal-to-noise ratio, which makes the removal
  of the telluric CO absorption challenging. The systematic error
brought in by the telluric correction is magnified by the
  subtraction of the outburst model. The broad line emission
is less subject to such problems in the measurements of the
equivalent widths. The equivalent width of  the narrow
line component looks constant in time.

\subsubsection{Disk Wind}

The model spectra of the broad line emission systematically
overestimate the flux of v=1-0 lines in the blue shoulder, and
underestimate the red.  After subtracting the spectral model,
therefore, the v=1-0 lines systematically show a blue depression
between $-$40 and $-$100~km~s$^{-1}$ with excess emission at the
red shoulder at $+$80~km~s$^{-1}$, which is similar to a P~Cyg
profile overlaid on the sharp quiescent emission (Fig.~\ref{dp},
right). This blueshifted absorption is more prominent toward the
later epoch. The absorption lines can be noted at $P$(21) to
$P$(26) on August 21, implying that the absorbing cloud is warm
($T_{\rm x}\gg$100~K). The absorption is not likely
photospheric, although EX~Lupi is an M-type star in its
quiescence, as the amount of blueshift is too large. The
emergence of blue shifted absorption is also observed in
infrared CO spectra of V1647~Ori toward the end of an outburst
\citep{bri07}. We infer that the blue absorption in the EX~Lupi
spectra also represents the disk wind, similarly to the case of
V1647~Ori.

\begin{figure}
\begin{center}
\includegraphics[height=0.52\textheight,angle=0]{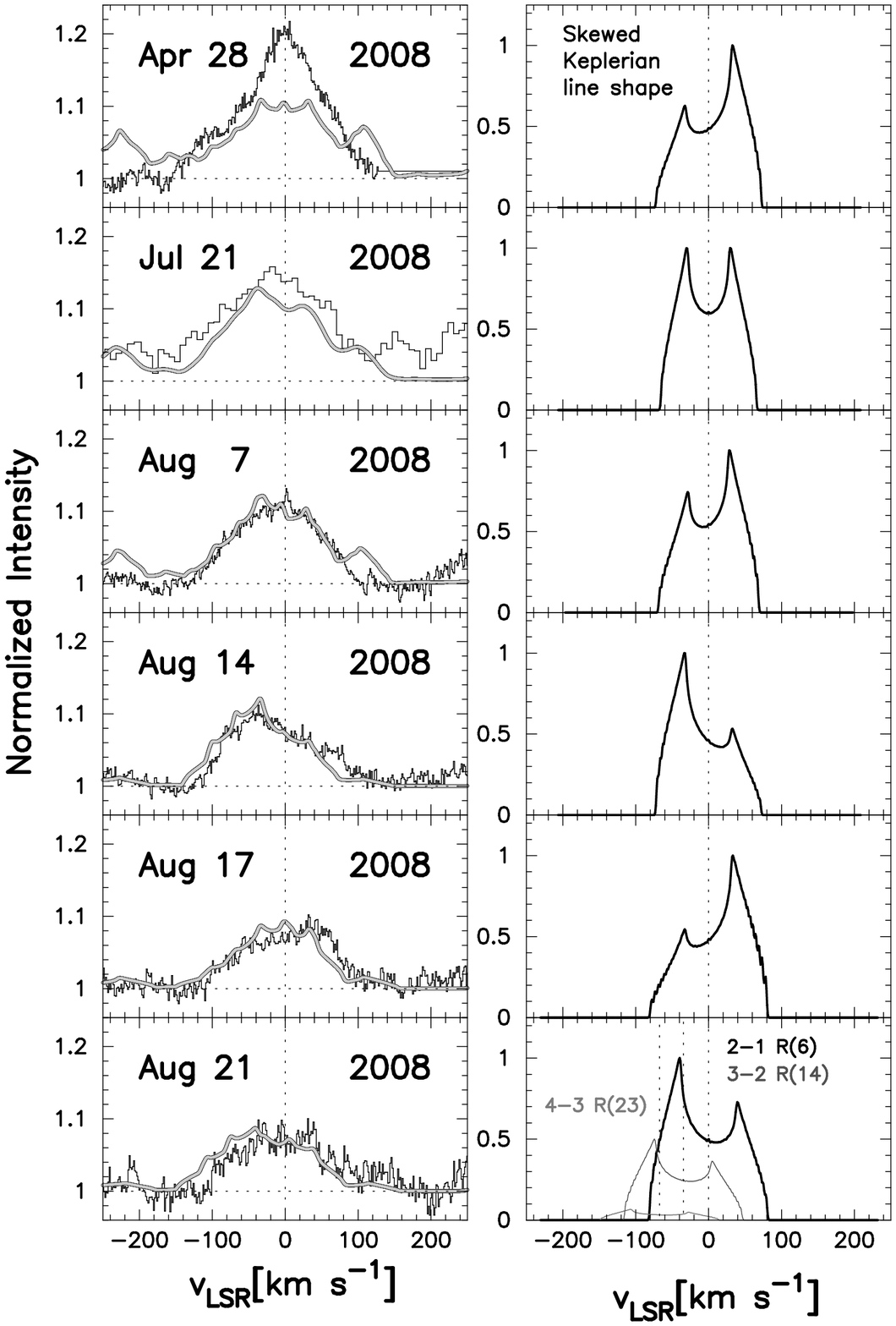}
\end{center}
\caption{Line profile of the $R$(6) v=2-1 transition at
  4.6675~$\mu$m. 
  The systemic velocity of EX~Lupi is close to 0~km~s$^{-1}$
  \citep{her03}. The model spectra of the outburst with $T_{\rm
    x}=4500$~K are overlaid in gray lines. There are minor
  contributions from the higher vibrational transitions of
  $R$(14) v=3-2 and $R$(23) v=4-3 as shown in the right bottom
  panel on 21 August. The observed line profiles cannot be
  reproduced without introducing the asymmetry. The asymmetry
  flips three times in August, which is consistent with the
  orbital period at 0.04--0.06~AU. \label{as}}
\end{figure}

\subsection{Heating Mechanism}

The excitation mechanism of the CO vibrational band is
  somewhat ambiguous. The elevated accretion energy can be
  directly converted to the kinetic energy of the gas through
  the viscous heating. In this case, the broad line emission
  that we locate at 0.04--0.4~AU traces the region where the
  accretion rate is high. On the other hand, the connection
  between the CO emission and the outburst could be
  indirect. The radiation from the central star, either from the
  hot accretion funnels or the footprints of the magnetic field
  lines, is first raised by the high disk accretion. The gas in
  the disk is heated afterwards by the enhanced irradiation.
  The primary difference of the two mechanisms is the vertical
  structure of the temperature that the heating imposes. In the
  case that the gas is viscously heated by the accretion, the
  disk is hottest near the mid-plane, and gradually cools down
  toward the surface. The emergent spectrum should be in
  absorption, as is the case for FUors \citep{har04}. When the
  irradiative heating is dominant, the temperature gradient is
  inverse with emission lines from the upper layer of the disk
  surface \citep{cal91a}.

  The irradiative heating has a few advantages to account for
  the CO line emission of EX~Lupi. First, most of the infrared
  CO emission lines observed today are thought to come from the
  hot surface of a disk, irradiatively heated
  \citep[e.g.][]{naj03}. This is fully consistent with the line
  flux of the narrow line emission closely following the
  continuum level. A critical problem of the direct heating of
  gas by the accretion is the cooling time of the outburst. The
  duration of the EX~Lupi outburst as is seen in the visible
  light curve, is roughly 8 months, from January to August
  2008. The viscous time scale on which the gas accretes on the
  star is given by
\[\frac{1}{t_{\rm vis}} = 3 \pi \, \alpha (H/r)^2 \cdot \frac{1}{t_{\rm orbit}},\]
for standard $\alpha$ prescription of an accretion disk. The
viscous time scale at 0.04~AU is 150 to 600 days, depending on
$H/r$ from 0.05 to 0.025 at the radius, even for the fastest
case with $\alpha$ being unity. The observed time scale of the
outburst is uncomfortably short compared to the most
optimistic case \citep{juh10}.

\begin{figure}
\begin{center}
\includegraphics[height=0.34\textheight,angle=0]{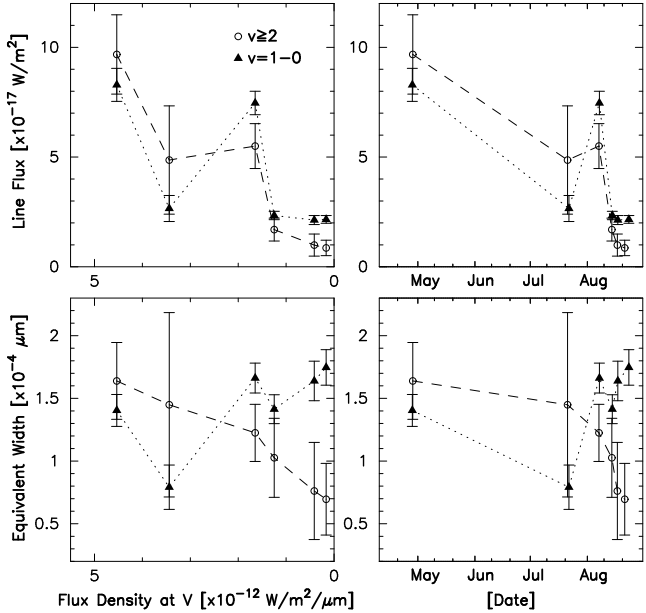}
\end{center}
\caption{Top: (Left) Line flux of the emission lines from
  vibrationally excited states (v$\ge$2, open circles), and the
  lowest transition v=1-0 (triangles) after subtraction of the
  outburst model of $T_{\rm x}=4500$~K shown as functions of
  the continuum flux of EX~Lupi at the visible wavelength.  The
  line intensity is integrated between $-$120 and
  $+$120~km~s$^{-1}$ for the broad line component.  The
  composite line profiles used in the calculations are shown in
  Fig.~\ref{dp}.  (Right) Same with the left panel, but shown
  with the date in abscissa.  Bottom: Same with the top panels
  but for the equivalent widths.  The line flux of the v$\ge$2
  and v=1-0 lines decrease with the time and the visual
  brightness of the star, but with different speed.  The
  difference is clearly seen in the plots of the equivalent
  widths.  The equivalent widths of the broader, highly excited
  lines decays quickly with the optical brightness of the star,
  while the equivalent widths of the v=1-0 narrow line emission
  stays constant during the outburst. \label{ew}}
\end{figure}

On the other hand, some features of the EX~Lupi outburst are not
fully explained by the irradiative heating either. The line flux of
both broad and narrow components decline with the visual
magnitude and the continuum flux at 5~$\mu$m (Fig.~\ref{ew},
upper panels), but on the slightly different decay rates. The
narrow line emission is diminished by the factor of 4 from April
to August (Table~\ref{t4}), which is the same factor with the
decline of 5~$\mu$m continuum flux, while the broad line
emission is decreased faster by the factor of 10.

In order to contrast the decay rates, the equivalent widths are
calculated in the same manner with the line flux. While the
equivalent width of the narrow line emission maintains the
level of 1.5$\times$10$^{-4}$~$\mu$m at the end of the outburst
(Fig~\ref{ew}, bottom right), where the visible magnitude of
EX~Lupi returned to $V=$12--13~mag, comparable to the level of
the pre-outburst brightness in 2007 (Fig.~\ref{lc}), the
equivalent width of the broad line emission declines, keeping
pace with the continuum flux at the visual wavelength. The
differential time scale of the two line components is not
accounted for, if the broad and the narrow line emissions are a
single disk component radiatively heated in the identical manner.

\begin{figure*}
\begin{center}
\includegraphics[height=0.2\textheight,angle=0]{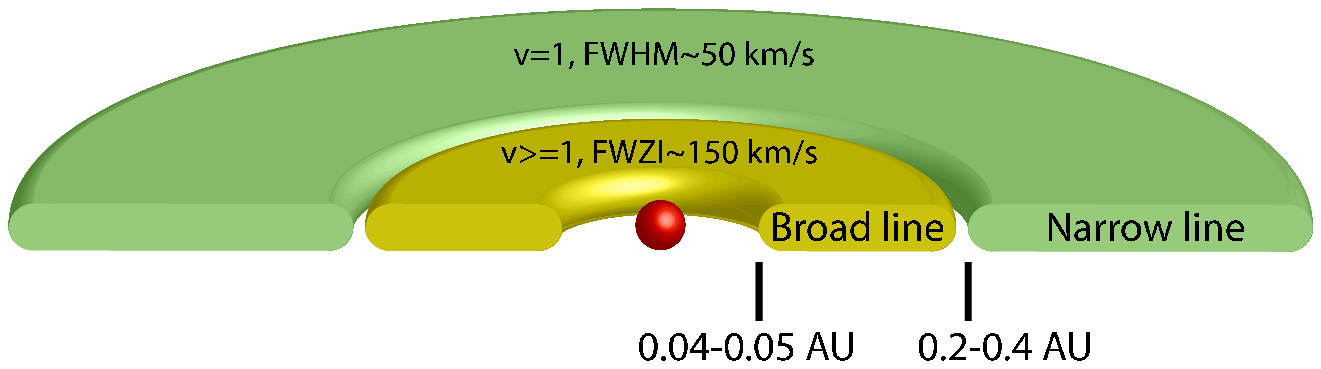}
\end{center}
\caption{Schematic representation of the broad and the narrow 
line emitting regions discussed in the text. 
  \label{car}} 
\end{figure*}

In addition, the excitation state and the kinematics are distinct
in two emission components, with no smooth transition between
them. The characteristic radius where the narrow line emission
arises from is 0.4~AU, and is constant as seen in the stable
line width. The radius is immediately outside of the broad line
emitting region at 0.04--0.4~AU (Fig.~\ref{car}).  The location
of this transition zone is close to the inner-rim of the dust
disk found by the SED modeling by \citet{juh10}. With the
vibrational excitation state upto v=6, and the faster decay of
the line flux than the radiatively heated narrow line component,
the broad line emission apparently traces the outburst event
more directly.

\setlength{\tabcolsep}{5pt}
\begin{deluxetable*}{lcccccc}
\tablecolumns{7} 
\tablewidth{0pc} 
\tablecaption{Line flux and equivalent width.\label{t4}}

\tablehead{
\colhead{Epoch}&
\colhead{JD}&
\colhead{$F_\lambda(V)$ }&
\multicolumn{2}{c}{Line flux}&
\multicolumn{2}{c}{Equivalent width}\\
\colhead{}&
\colhead{}&
\colhead{[$\times$10$^{-12}$ Wm$^{-2}$$\mu$m$^{-1}$]}&
\multicolumn{2}{c}{[$\times$10$^{-17}$ Wm$^{-2}$]}&
\multicolumn{2}{c}{[$\times$10$^{-4}$ $\mu$m]}\\
\colhead{}&
\colhead{}&
\colhead{}&
\colhead{v$\ge$2\tablenotemark{a}}&
\colhead{v$=$1-0\tablenotemark{b}}&
\colhead{v$\ge$2}&
\colhead{v$=$1-0}}

\startdata
28 Apr 2008  &2454585 & 4.5 &  9.7$\pm$ 1.8 &  8.3$\pm$ 0.8 & 1.6$\pm$ 0.3 & 1.4$\pm$ 0.1\\
21 Jul 2008   &2454669 & 3.4 &  4.9$\pm$ 2.5 &  2.7$\pm$ 0.6 & 1.4$\pm$ 0.7 & 0.8$\pm$ 0.2\\
~7 Aug 2008  &2454686 & 1.6 &  5.5$\pm$ 1.0 &  7.5$\pm$ 0.5 & 1.2$\pm$ 0.2 & 1.7$\pm$ 0.1\\
14 Aug 2008  &2454693 & 1.2&  1.7$\pm$ 0.5 &  2.3$\pm$ 0.2 & 1.0$\pm$ 0.3 & 1.4$\pm$ 0.1\\
17 Aug 2008  &2454696 & 0.4&  1.0$\pm$ 0.5 &  2.1$\pm$ 0.2 & 0.8$\pm$ 0.4 & 1.6$\pm$ 0.2\\
21 Aug 2008  &2454700 & 0.2&  0.9$\pm$ 0.4 &  2.2$\pm$ 0.2 & 0.7$\pm$ 0.3 & 1.7$\pm$ 0.1\\

\enddata

\tablenotetext{a}{Integrated over $\pm$120~km~s$^{-1}$.}
\tablenotetext{b}{Integrated over $\pm$40~km~s$^{-1}$, after the spectral model with $T_{\rm x}=4500$~K is subtracted.}

\end{deluxetable*}

The heating mechanism of similar outburst variables, FUors, is
most likely the elevated accretion near the disk-midplane, as
the broad and double-peak CO lines in absorption testify
\citep[e.g.][]{har04}. To the contrary, the CO vibrational lines
of EX~Lupi are observed in emission in both overtone
\citep{kos10} and fundamental bands. The critical difference
that might explain the emergent spectra of the two classes of
the outburst variables is the optical depth of the disk. The
broad line emitting region of EX~Lupi is likely free of dust as
is discussed above. The SED analysis by \citet{juh10} also
  excludes the optically thick continuum emission either by gas
  or by dust grains within 0.3~AU during the outburst. Moreover,
  the accretion rate during the outburst is a few orders
  magnitude smaller in EX~Lupi ($\sim 10^{-7}M_\odot$~yr$^{-1}$)
  than FUors ($\sim 10^{-4}M_\odot$~yr$^{-1}$), implying the
  lower disk surface density. The disk close to the midplane
  that is viscously heated by the accretion might be more
  exposed to the surface in the case of EX~Lupi, which possibly
  makes the CO vibrational band observed in emission.

\subsection{Origin of Outburst}

Let us extend our speculation to the trigger of the outburst,
taking the physical scale of it being traced by the broad line
emission. The physical size of the outburst on the scale of
0.2--0.4~AU argues against mechanisms that predict a global disk
instability. \citet{bon92} proposed that a binary system can
tidally disrupt the disks of each other in a close encounter,
and temporarily increase the accretion rate up to
10$^{-4}M_\odot$~yr$^{-1}$.  The tidal disruption by a companion
is attractive because it naturally explains the repetitive
nature of EXor outbursts. The binary fraction among low-mass
pre-main sequence stars is indeed high
\citep[e.g.][]{ghe93}. FU~Ori, the prototype FUor, has a close
companion \citep{wan04}, and there are even FUor-like binaries
where both of the binary components are FUor-like, which is
statistically unlikely if two are randomly paired
\citep{rei04}. The gravitational disruption by a binary
companion may work for the present case as a trigger of the disk
instability combined with other mechanisms, but most likely does
not explain the small radius of the high mass-accretion region
by itself. No visible \citep{ghe97,bai98} or spectroscopic
companion \citep{her07} to EX~Lupi has been identified as of
today.

\citet{vor06} argued that planetary cores -- which possibly have
already formed in a gravitationally unstable disk -- can trigger
an FUor outburst when they migrate inward and eventually fall
into the central star. Protoplanetary clumps form at
$>$10--50~AU away from the star. The migration from the outer
disk over this distance does not explain the discrete boundary
at 0.2--0.4~AU either. Their simulation ends at the inner radius
of 0.5~AU. Further investigation how the infalling a
protoplanetary clump behaves in the final 0.5~AU is awaited.

When mass inflow from the outer disk is faster than what the
inner disk can transport further in, the accretion slows down,
and piles up material at the boundary between the inner and 
the outer disk. The elevated surface density
triggers a local disk instability, and the accretion mode
switches to a higher state. The mass accretion could slow down
at a certain radius simply because the disk viscosity is too
low, or the viscosity is locally too low because of the
decoupling of the disk and the magnetic field, or if the disk
has a gap, or if it is magnetically truncated.

The disk viscosity is provided by the coupling of the magnetic
field and slightly ionized medium in the MHD instability model. Deep
inside the disk where the stellar radiation no longer penetrates, 
there is a low ionization region called the dead zone, where
the viscosity is close to zero. The incoming mass piles up there,
until the surface density becomes high enough to heat the disk
to $\approx$1000~K to restore the coupling with the magnetic
field.

A dead zone starts at 0.1~AU, and extends beyond 1~AU in a disk
around a solar mass star \citep{gam96}. The disk instability is
triggered at the coldest region of the dead zone near the outer
boundary \citep{arm01}. \citet{zhu07} constructed a radiation
transfer model of the disk around FU~Ori in outburst, and
calculated the size of the active region to be around 1~AU, which
naturally fits with MHD disk instability. The observed transition
region of EX~Lupi is a few times smaller than the typical size of
a dead zone. This may not be surprising, as the luminosity of
EX~Lupi in quiescence is small as well
[$L_\ast<0.5L_\odot$ in Gras-Vel\'azquez \& Ray (2005), 
$L_\ast=0.75L_\odot$ in Sipos et al. (2009)].  

\citet{wun05} discovered another type of instability near the
inner edge of the dead zone in their numerical model. When a small
perturbation is applied to the height of the dead zone, the mass
inflow locally increases while the outgoing mass decreases. The
height of the perturbed region grows, until it splits up from
the dead zone and triggers a ring instability, or the inner edge
of the dead zone starts oscillating radially \citep{wun06}. This
instability works at the correct physical scale of 0.1--0.2~AU with
small accretion rates $10^{-9}M_\odot$~yr$^{-1}$ which are appropriate 
for EX~Lupi. However, the predicted timescale of the outburst is
100~yrs or longer. This is close to the observed timescales of 
FUor outburst, but is too long for EX~Lupi ($\sim$1~yr).

When a star is strongly magnetized, the gas infall stops at the
radius where the Keplerian angular velocity is equal to the
stellar rotation, as the orbital motion of the gas is tied to
the star. The matter piles up at the co-rotation radius, until
the surface density becomes so high that the thermal pressure of
the gas overcomes the magnetic pressure \citep{dan10}. This
mechanism requires the presence of a strong magnetic
field. Among young eruptive stars, FU~Ori is the first where a
magnetic field was discovered in its disk \citep{don05}. There
are no attempts known to date, however, to measure the magnetic
field of EX~Lupi. The inner rim of the high accretion region we
measured (0.04--0.05~AU) translates to a period of 4--5 days
assuming Keplerian rotation. The typical rotation time for
T~Tauri stars is a few days to a week \citep[e.g.][]{rod09}. The
rotation period of EX~Lupi itself is however not known yet.

A self-regulated thermal instability theory assumes relatively
high, constant inflow of mass
\citep[$10^{-6}M_\odot$~yr$^{-1}$--$10^{-7}M_\odot$~yr$^{-1}$,
][]{cla89,bel94}. The constant mass transportation is hindered
at a certain radius, where the viscosity is simply too low to
drain the incoming mass in time. The material piles up until the
surface density becomes high enough to thermally ionize the
gas. Near the ionization front, the disk opacity increases with
temperature due to negative hydrogen ions adding continuum
opacity \citep{fau83}. Thermal instability is triggered by the
runaway local heating of the disk. The transition front
propagates outward until the instability is suppressed at the
radius where the viscous heating is no longer effective to keep
the disk warm. The radius at which the thermal instability is
turned off is a function of the mass accretion. It is 0.1~AU for a
fiducial star of 1~$M_\odot$ with 3~$R_\odot$ with a constant
mass accretion of $3\times 10^{-6} M_\odot$ \citep{bel94}.

Self-regulating thermal instability is well studied in the
parameter space suited to FUors. The mechanism seems to work
with accretion rates from $\dot{M}=10^{-7}M_\odot$~yr$^{-1}$ to
$\dot{M}=10^{-5}M_\odot$~yr$^{-1}$, and is able to reproduce the
observed flaring magnitude at optical wavelengths
($\approx$5~mag), the rise time to the maximum ($\sim$~1 year),
and the duration of the outburst (decades to a century). These
physical parameters are one order of magnitude smaller in the
case of EX~Lupi, with the quiescent phase accretion rate being
$\dot{M}=4\times 10^{-10}M_\odot$~yr$^{-1}$ \citep{sip09}, which
goes up to $\dot{M}=2\times 10^{-7}M_\odot$~yr$^{-1}$ in the
outburst \citep{juh10} with the rise time of a month, and
duration of about a year. It is yet to be seen if the
self-regulating thermal instability works for the small
accretion rate with no external triggers to cooperate.

An interesting possible triggering mechanism is an embedded
planet. When a protoplanet opens a gap, the mass transportation
stops at its outer edge. The mass is banked up until the thermal
instability is triggered \citep{lod04,cla96}. With the
additional degrees of freedom of the mass of the planet and its
location, this model could cover a wider parameter space, and
closely reproduce the observed properties of FUor outbursts. The
triggering via embedded planet makes the instability propagate
initially outside in, and qualitatively reduces the rise
time. The smaller the mass of the planet, the smaller the
outburst, triggered by the smaller accretion rate. It is still
to be seen though, if the mechanism works for EX~Lupi, as the
exact parameter space of EX~Lupi outburst is not covered in the
models by \citet{lod04}.

\subsection{Variation of Line Profile Asymmetry}

Nevertheless, a possible planet in the disk is interesting in
connection with the temporal variation of the line symmetry in
CO emission spectra (Fig.~\ref{as}). The line asymmetry
indicates that the receding side of the disk was brighter on
August 7, but by August 14 the approaching side became
brighter. The receding part is again brighter 3 days later,
which again moved to the other side in 4 days. The modulation of
the line profile on the timescale of 3--4 days was known in the
visible spectra of FU~Ori and V1057~Cyg. \citet{her03} discussed
two possible origins of the variation, that the modulation comes
from the stellar spot and that it comes from the hot spot in the
circumstellar disk, which may or may not be related to the hot
accretion columns rising out of the disk. In the case of
V1057~Cyg, the line modulation is not accompanied by
photospheric variability, which makes the stellar spot
hypothesis difficult. On the other hand, the line modulation in
V1057~Cyg is stable over 3 yrs, which also makes the hot spot on
a differentially rotating disk unlikely.  The question of
whether the hot spot resides on the photosphere or on the disk
is therefore still open.

The period of the spectroscopic modulation of EX~Lupi is hard to
constrain with only six epochs of observations. If we simply
take 7 and 3.5~days as half the period of the orbit in the first
and the second half of August, the locations of the hot spot are
0.06 and 0.04~AU, respectively. This is within the span of the
active disk that we derived from the outburst model, with the
latter location close to the inner edge. Further spectroscopic
monitoring with finer temporal sampling should be done in order
to tell whether this represents real migration of a hot spot in
the disk, as well as the theoretical study on the fate of
protoplanteary clumps that fall into the star \citep{vor06}.  A
more general discussion on the line asymmetry caused by a planet
embedded in the disk is found in \citet{reg10}.

\section{Summary}

Here is the summary of our results, and possible directions for
future investigations.

\begin{itemize}

\item We reported the results of a spectroscopic monitoring
  campaign of EX~Lupi in the CO vibrational fundamental band at
  4.6--5.0~$\mu$m. The observations covered six epochs in the
  outburst, which started in early 2008.

\item We detected high temperature CO gas in the emission-line
  spectra, that apparently consist of two distinct
  components. The equivalent width of the broad line emission
  (FWZI$\sim 150$~km~s$^{-1}$) decays with the visual brightness
  of the star. The gas is vibrationally hot with the highly
  excited lines detected up to v=6.

\item The line profiles of the broad line component are double
  peaked, indicating that the gas is in a circumstellar disk. We
  used a simple slab model to locate the hot gas to the annulus
  0.04--0.4~AU. There is strong asymmetry in the line profiles
  that varies with time on the order of days to week, implying
  the presence of a hot spot on the disk.

\item There is a second component of the line emission near the
  systemic velocity of EX~Lupi with much narrower line profiles
  (FWHM$\approx$50~km~s$^{-1}$). The narrow line emission is
  vibrationally cool; it exclusively consists of v=1-0
  transitions.  The line width indicates that the characteristic
  radius of the emitting gas is 0.4~AU, larger than the radius
    where the broad line emission arises. The equivalent widths
  of the narrow lines are constant over the whole observing
  period.

\item An additional disk wind is found as absorption lines
  centered at $-$80~km~s$^{-1}$ toward the end of the
  outburst. A similar disk wind was observed previously in
  V1647~Ori.

\item The two emission components are distinct in the decay
  rate, the excitation state, and the kinematics. If the broad
  line emission traces the high accretion region, its compact
  size favors local disk instabilities where a disk goes through
  bimodal accretion. Major local disk instability theories, such
  as thermal instability, magnetohydrodynamical instability,
  magnetospheric truncation of a disk partially fit to the
  profile of the outburst, albeit our data did not allow to
  distinguish between these theories.  The origin of the
  outburst involving global disk instabilities are unlikely.

\end{itemize}

Most of the disk instability theories today have been tested
against FUor outbursts, but little theoretical work has been 
done for EXors. It is not clear at
the moment if the missing proper model for EX~Lupi outbursts is
due to an intrinsic difficulty in reproducing such small and
frequent outbursts with low accretion rate, or is due to a lack of
exploration in the parameter space that fits to EXor outbursts.
Our attempts to remedy this are also hindered by missing fundamental
parameters of EX~Lupi. The period of the stellar rotation and
magnetism, as well as an accurate simultaneous photometric 
calibration of the spectra at the time of the observation would 
help to improve the modeling of the outburst.
 
\acknowledgments {We appreciate the anonymous referee for the
  constructive criticisms to improved the manuscript. We thank
  all the staff and crew of the {\it VLT} and Subaru for their
  valuable assistance in obtaining the data. M.G. thanks Takeshi
  Oka for patient enlightening of her in thermodynamics of
  CO. M.G. also thanks Bringfried Stecklum and Andreas Seifahrt
  for instructive discussion and encouragement to finish the
  paper. Zs.R. was partly supported by M\"OB/DAAD-841 Mobility
  Grant. A.C acknowledge support from a Swiss National Science
  Foundation grant (PP002--110504). P.\'A. acknowledges support
  from the grant OTKA K62304 of the Hungarian Scientific
  Research Fund. The research of \'A.K. is supported by the
  Netherlands Organization for Scientific Research.
  A.S.A. acknowledges support from the Deutsche
  Forschungsgemeinschaft (DFG) grant SI 1486/1-1.  We appreciate
  the hospitality of Chilean and Hawaiian community that made
  the research presented here possible.}

\end{document}